\documentclass[12pt]{iopart} 
\usepackage{iopams}
\usepackage{graphicx} 
\usepackage{epstopdf}
\usepackage{url}
\usepackage{hyperref}
\usepackage[usenames,dvipsnames]{color}
\begin{document}

   \title{An Analysis of Characteristics in Non-Linear Massive Gravity}

   \author{Keisuke Izumi$^1$ and Yen Chin Ong$^{1,2}$}
    \address{$^1$ Leung Center for Cosmology and Particle Astrophysics,\\ National Taiwan University, Taipei, Taiwan 10617.}
		\address{$^2$ Graduate Institute of Astrophysics, National Taiwan University,\\ Taipei, Taiwan 10617.}
    \eads{\mailto{izumi@phys.ntu.edu.tw}, \mailto{ongyenchin@member.ams.org}}

\date{\today}

\begin{abstract}
We study the Cauchy problem in a special case of non-linear massive gravity. Despite being ghost-free, it has recently been argued that the theory is inherently problematic due to the existence of superluminal shock waves. Furthermore it is claimed that acausal characteristic can arise for \emph{any} choice of background. In order to further understand the causal structure of the theory, we carefully perform a detailed analysis of the characteristic equations and show that the theory does admit a well-posed Cauchy problem, i.e., there exists hypersurfaces that are \emph{not} characteristic hypersurface. Puzzles remain regarding the existence of a superluminal propagating mode in both the minimal ghost-free theory that we analyzed, as well as in the full non-linear massive gravity. That is, our result should not be taken as any indication of the healthiness of the theory. We also give a detailed review of Cauchy-Kovalevskaya theorem and its application in the Appendix, which should be useful for investigating causal structures of other theories of gravity.
\end{abstract}

\maketitle
\section{Introduction}

The question as to whether graviton can have a mass is an old one. A recent review of the literature can be found in Ref.\cite{Hinterbichler}. Here we give a very brief review of the main progress and major difficulties encountered in constructing a healthy massive gravity theory. 

It is well-known that Fierz and Pauli constructed a theory of a free massive spin-2 particle back in 1939 \cite{FP}, in which the quadratic form of the action can be uniquely fixed as linearized general relativity (GR) with the Fierz-Pauli mass term, by imposing the tachyon-free and ghost-free conditions \cite{Nieuwenhuizen}. However, it was discovered in 1970 that this linear massive spin-2 theory suffers from the van Dam-Veltman-Zakharov (vDVZ) discontinuity \cite{vDV,Z}, with a massless limit that does not recover that of linearized  general relativity. In particular, the prediction of light-bending near massive objects was off by 25\% and is therefore completely ruled out by observations. The \emph{Vainshtein mechanism} was proposed in 1972, in which a non-linear effect is introduced to force the massless limit to recover GR \cite{Vainshtein, BDZ}. However in the same year it was shown that the exact non-linearity introduced gives rise to the \emph{Boulware-Deser ghost} or BD ghost for short \cite{BD}, where the mass of the ghost mode is typically the same as that of the graviton, which means the ghost mode cannot be ignored (e.g., by pushing to the Planck scale to be delt with by a quantum theory of gravity \cite{Pisin}) since its mass is expected to be \emph{small}. The BD ghost is essentially due to an extra degree of freedom reinstated by the non-linearity, beyond the linearized 5 degrees of freedom. The interest in massive gravity was revived recently due to the construction of a presumably consistent non-linear massive gravity without the BD ghost \cite{dRGT1, dRGT2}. The absence of the BD ghost was checked at low order perturbative analysis level in the metric perturbation approach \cite{dRGT3, dRGT4},  
and subsequently even at fully non-linear orders \cite{HR, HR1, HR2, HSMS}.
It is interesting to note that, as pointed out in \cite{DW}, Wess-Zumino already constructed a theory without the 6th degree of freedom and thus free of the BD ghost in 1970 \cite{WZ}.
However, it should be pointed out this is a realization by hindsight: Wess-Zumino had simply presented a number of massive gravity models  the linear Fierz-Pauli theory, without calculating the number of degrees of freedom, and without considering the issue of ghosts. We shall henceforth refer to such presumably consistent theories as non-linear massive gravity in general. In non-linear massive gravity, in addition to the dynamical metric $g_{\mu\nu}$, there is also fixed background with non-dynamical metric  $\bar g_{\mu\nu}$, also known as the \emph{fiducial metric}.

Recently, the characteristic equations of a special case of non-linear massive gravity, essentially one of the original model of Wess-Zumino with 5 degrees of freedom, were analyzed by Deser and Waldron \cite{DW} (see also the newer work \cite{DW2}). It was argued that the model admits superluminal shock wave solutions and therefore is acausal, and in addition, the superluminal shocks are only absent in the special case where the contortion term vanishes. Furthermore, ironically this originates from the very constraint that removes the BD ghost. In the same work, it is mentioned that this fatal problem arises in \emph{any} choice of background. This is an intriguing claim, which if true, is a rather strong one. After all, we do know that at least some physics \emph{does} depend on the choice of background. For example, a FLRW (Friedmann-Lema\^itre-Robertson-Walker) background does not admit the flat or closed cosmology\footnote{This is also caused by the very constraint that removes the BD ghost.} \cite{massivecosmo}, however more general fiducial metrics do allow FLRW universes of all signs of spatial curvature \cite{GLM}. Motivated by the interesting work of Ref.{\cite{DW}}, we perform a careful analysis of the characteristic equations to further understand the causal structure of non-linear massive gravity. 

In section \ref{section2}, we introduce the necessary set-up. Section \ref{section3} contains the main analysis of the characteristic equations, while section \ref{section4} gives a concrete example to show a well-posed Cauchy problem in the theory. We conclude with a discussion on the existence of superluminal propagation in non-linear massive gravity in section \ref{section5}. 
The method of characteristic employed in our work is reviewed in detail in \ref{app}.

\section{Non-Linear Massive Gravity Theory}
\label{section2}

As mentioned in the introduction, in non-linear massive gravity, we need to first choose and fix a background with a non-dynamical reference metric $\bar g_{\mu\nu}$, in addition to the physical metric $g_{\mu\nu}$. 
For convenience of our analysis, we will work in the tetrad formalism \cite{Chamseddine:2011mu, Chamseddine:2011bu, Hinterbichler:2012cn, Hassan:2012wt, Nomura:2012xr}, following the work of Ref.\cite{DW}%
\footnote{
 The massive gravity theory in the tetrad formalism is, at least in classical physics, the same as the non-linear massive gravity proposed in Refs.\cite{HR, HSMS, dRGT3, dRGT4}.
}. 
We denote the background tetrad by $f^a_{~\mu}$, and the dynamical tetrad by $e^a_{~\mu}$. Here Greek letters correspond to spacetime indices, while Latin letters label tetrad vectors. The theory is also referred to as the two-metric ``$f$-$g$'' theory in Ref.\cite{DW}. However, since there is a risk of confusion with bimetric theory, in which both metrics are dynamical, we will refer to this model as the \emph{minimal ghost-free model} instead, since this is the simplest of the ghost-free models in the formulation of Ref.\cite{HR3}.

The Levi-Civita connection $\omega(e)_\mu{}^m_{~n}$ that corresponds to the dynamical tetrad is torsion-less. However, the same connection employed as the connection for the background tetrad has nonzero torsion in general. The difference of the two connections is the contortion tensor \footnote{We would like to remark that the proper term is \emph{contortion} instead of ``contorsion''. See e.g., \cite{Hehl}.}:
\begin{eqnarray}
K_{\mu}{}^m_{~n} := \omega(e)_\mu{}^m_{~n} - \omega(f)_\mu{}^m_{~n}.
\end{eqnarray}
The background metric $\bar g_{\mu\nu}$ and the dynamical metric $g_{\mu\nu}$ are related to the background tetrad and the dynamical tetrad, respectively, by
\begin{eqnarray}
f^a_{~\mu}f^b_{~\nu}\eta_{ab}=\bar g_{\mu\nu} \qquad \mbox{and} \qquad
e^a_{~\mu}e^b_{~\nu}\eta_{ab}=g_{\mu\nu}.
\end{eqnarray}
Raising and lowering of the Greek indices are performed via the dynamical metric, i.e., 
\begin{eqnarray}
V^\mu=V_\nu g^{\nu\mu}, \quad \mbox{and} \quad V_\mu=V^\nu g_{\nu\mu}.
\end{eqnarray}

The action of the non-linear massive gravity theory is defined as \cite{HR3}
\begin{eqnarray}
S= \frac{1}{2\kappa} \int d^4x \ e \left(  R + 2 \sum_{n=0}^{4} \alpha_n {\cal L}_n \right),
\end{eqnarray}
in the unit $c=1$, 
where the $\alpha_n$'s are constant parameters of the theory and the exact expressions of the ${\cal L}_n$'s are
\begin{eqnarray}
&&{\cal L}_0=1,\\
&&{\cal L}_1=f,\\
&&{\cal L}_2=f^2-f^\mu_{~\nu} f^\nu_{~\mu},\\
&&{\cal L}_3= f^3-3ff^\mu_{~\nu} f^\nu_{~\mu}+2f^\mu_{~\nu} f^\nu_{~\lambda}f^\lambda_{~\mu},\\
&&{\cal L}_4=f^4-6f^2f^\mu_{~\nu} f^\nu_{~\mu}+8ff^\mu_{~\nu} f^\nu_{~\lambda}f^\lambda_{~\mu}+3(f^\mu_{~\nu} f^\nu_{~\mu})^2-6f^\mu_{~\nu} f^\nu_{~\lambda}f^\lambda_{~\delta} f^\delta_{~\mu}.
\end{eqnarray}
Here, $f_{\mu\nu}=e^a_{~\mu}f^b_{~\nu}\eta_{ab}$ and $f=f^\mu_{~\mu}$.

For simplicity, we consider the case where $\alpha_2=\alpha_3=\alpha_4=0$. 
The equation of motion can be obtained by the variation with respect to the dynamical tetrad $e^a_{~\mu}$:
\begin{eqnarray}
{\cal G}_{\mu\nu}:=G_{\mu\nu} - \alpha_0 g_{\mu\nu} + \alpha_1(f_{\nu\mu}-fg_{\mu\nu})=0, \label{EOM}
\end{eqnarray}
where $G_{\mu\nu}$ is the Einstein tensor.
Here, in order to have a simpler form of equation of motion, we have acted upon the expression obtained after taking variation with respect to $e^a_{~\mu}$ by $-e_{a\mu}g_{\alpha\nu}$. We emphasize that
$f_{\mu\nu}$ has all of the information of the dynamical tetrad $e^a_{~\mu}$, and thus we analyze the dynamics of $f_{\mu\nu}$ instead of that of $e^a_{~\mu}$.
From the antisymmetric components of Eq.(\ref{EOM}), we find that $f_{\mu\nu}$ is a symmetric tensor and thus has ten degrees of freedom. 

Equation (\ref{EOM}) includes not only dynamical equations but also five constraint equations, which we shall now derive.
The first two terms in Eq.(\ref{EOM}) are divergence free, and thus, taking the divergence of Eq.(\ref{EOM}) gives four constraint equations \cite{DMZ}:
\begin{eqnarray}
{\cal C}_\mu:=\nabla^{\nu} f_{\mu\nu}+\nabla_\mu f=0. \label{vconst}
\end{eqnarray}
Another constraint equation \cite{HR2, HSMS} stems from the nontrivial combination of the trace of Eq.(\ref{EOM}) and  Eq.(\ref{vconst}); it is
\begin{eqnarray}
{\cal C}:=2 \nabla_\mu\left( l^{\mu\nu} {\cal C}_\nu\right) - {\cal G}^\mu_{~\mu}=0,
\label{sconst}
\end{eqnarray}
where $l^{\mu\nu}=l^\mu_{~a}e^{a\nu}$ and $l^{\mu}_{~a}$ is the inverse of the background tetrad $f^a_{~\mu}$. 
From the symmetry of the indices of $f_{\mu\nu}$, we can infer that the indices of $l^{\mu\nu}$, which is simply the inverse of $f_{\mu\nu}$, are also symmetric.
In the combination (\ref{sconst}), the terms involving second order derivatives of $f_{\mu\nu}$ are canceled out, and it becomes a constraint equation. 
To sum up, the ten symmetric components of Eq.(\ref{EOM}) include the five constraint equations given by Eq.(\ref{vconst}) and Eq.(\ref{sconst}), and five dynamical components containing the second order derivatives of $f_{\mu\nu}$. 

\section{Characteristic in Non-Linear Massive Gravity}
\label{section3}

Suppose we have a hypersurface $\Sigma$ and $\xi^\mu$ is the unit normal vector to $\Sigma$. 
Here, we consider the case where $\xi^\mu$ is timelike. 
We use Gaussian normal coordinate near the hypersurface $\Sigma$,
\begin{eqnarray}
ds^2=-\left(dx^0\right)^2 + h_{ij}dx^idx^j, \quad \mbox{with}
\quad \left(\frac{\partial}{\partial x^0}\right)^\mu= \xi^\mu.
\label{metric}
\end{eqnarray}
In Eq.(\ref{EOM}), all second derivatives of $f_{\mu\nu}$ are of the form $\partial_\mu \partial_{[\nu} f_{\alpha]\beta}$ where the bracket $[\cdot\cdot]$ represents antisymmetrization, i.e., 
\begin{eqnarray}
T_{\alpha \cdots [\beta \gamma]\cdots}:= \frac{1}{2}\left(
T_{\alpha \cdots \beta \gamma\cdots}-T_{\alpha \cdots \gamma\beta\cdots}\right).
\end{eqnarray}
Having in mind the Cauchy-Kovalevskaya theorem \cite{C,K} (See Appendix A for a detailed discussion), we shall introduce new variables  
\begin{eqnarray}
M_{\alpha\beta\gamma}:=\partial_{[\alpha} f_{\beta]\gamma}.\label{defM}
\end{eqnarray}
Since $M_{\alpha\beta\gamma}$ is antisymmetric with respect to $\alpha$ and $\beta$, we naively expect that $M_{\alpha\beta\gamma}$ has 24 degrees of freedom. There are, however, 4 constraints for $M_{\alpha\beta\gamma}$, namely,
\begin{eqnarray}
\epsilon^{\mu\alpha\beta\gamma}M_{\alpha\beta\gamma}=0, \label{constM}
\end{eqnarray}
where $\epsilon^{\mu\alpha\beta\gamma}$ is the Levi-Civita symbol. This fact can be
deduced from the definition (\ref{defM}) of $M_{\alpha\beta\gamma}$. Therefore the actual number of degrees of freedom of $M_{\alpha\beta\gamma}$ is only 20.  
Under 3+1 decomposition of spacetime based on the metric (\ref{metric}), 
all non-vanishing components can be written as 
\begin{eqnarray}
&&M_{0i0}=-M_{i00}=\frac{1}{2}\left(\partial_0 f_{i0}-\partial_if_{00}\right),\label{0i0}\\
&&M_{0ij}=-M_{i0j}=\frac{1}{2}\left(\partial_0 f_{ij}-\partial_if_{0j}\right),\label{0ij}\\
&&M_{ij0}=-M_{ji0}=\frac{1}{2}\left(\partial_i f_{j0}-\partial_jf_{i0}\right),\label{ij0}\\
&&M_{ijk}=-M_{jik}=\frac{1}{2}\left(\partial_i f_{jk}-\partial_jf_{ik}\right).\label{ijk}
\end{eqnarray}
Using the four constraints (\ref{constM}) for $M_{\alpha\beta\gamma}$, 
we can express three of $M_{0ij}$ and one of $M_{ijk}$ by the other $M_{\alpha\beta\gamma}$'s.
Thus, after careful comparison it can be deduced that $M_{0i0}$, $M_{0ij}$, $M_{ij0}$ and $M_{ijk}$ have three, six, three and eight degrees of freedom respectively. 
In addition to the four constraints (\ref{constM}), we can derive other relations among the $M_{\alpha\beta\gamma}$'s. They are:
\begin{eqnarray}
&&\partial_0 M_{ij0}= \partial_i M_{0j0}-\partial_j M_{0i0},\label{0ij0}\\
&&\partial_0 M_{ijk}= \partial_i M_{0jk}-\partial_j M_{0ik}.\label{0ijk}
\end{eqnarray}

Equations (\ref{0i0}), (\ref{0ij}), (\ref{0ij0}) and (\ref{0ijk}) can be regarded as the time-evolution equations for $f_{i0}$, $f_{ij}$, $M_{ij0}$ and $M_{ijk}$ respectively. 
All coefficients of the terms with a $x^0$-derivative in these equations are obviously nonzero. 
Therefore, the hypersurface $\Sigma$ is \emph{not} a characteristic surface for $f_{i0}$, $f_{ij}$, $M_{ij0}$ and $M_{ijk}$. 

Now, in order for the time evolution to be well-defined and solvable, we need to find appropriate equations to fix the time derivatives of the ten variables: $f_{00}$, $M_{0i0}$ and $M_{0ij}$. 
We have the ten components of the equation of motion (\ref{EOM}), and so, should expect that the time derivatives of these ten variables can be obtained from them. 

Having introduced $M_{\alpha\beta\gamma}$, 
 all second order derivatives of $f_{\mu\nu}$ in the equation of motion (\ref{EOM}) can be written in terms of the first order derivatives of $M_{\alpha\beta\gamma}$.
Then, the equation of motion (\ref{EOM}) becomes 
\begin{eqnarray}\label{Gmunu}
&2\left( g^{\alpha\gamma}l^{\delta}_{~(\mu} \delta^\beta_{~\nu)}
+l^{\beta\delta}\delta^\alpha_{~(\mu}\delta^\gamma_{~\nu)} 
-g^{\alpha\gamma}l^{\beta\delta}g_{\mu\nu}\right) \partial_\alpha M_{\beta\gamma\delta} +\bar F_{\mu\nu}\nonumber\\
&\quad
+l^\gamma_{~b}\left(\partial_\delta f^b_{~\lambda}\right)
\biggl(l^{\lambda\beta}\delta^\alpha_{~(\mu}\delta^\delta_{~\nu)}
-l^{\lambda\delta}\delta^\alpha_{~(\mu}\delta^\beta_{~\nu)}
+g^{\alpha\delta}l^\lambda_{~(\mu}\delta^\beta_{~\nu)}
\nonumber\\
&\qquad\qquad\qquad\qquad
-g^{\alpha\beta}l^\lambda_{~(\mu}\delta^\delta_{~\nu)}
-g^{\alpha\delta}l^{\lambda\beta}g_{\mu\nu}
+g^{\alpha\beta}l^{\lambda\delta}g_{\mu\nu} \biggr) \partial_\alpha f_{\beta\gamma}
\nonumber\\
&\quad
+N_{\delta\epsilon\lambda}\biggl(
2g^{\alpha\gamma}l^{\beta\delta}l^\lambda_{(\mu}\delta^\epsilon_{~\nu)}
-2l^{\beta\epsilon}l^{\lambda\delta}\delta^\alpha_{~(\mu}\delta^\gamma_{~\nu)}
+2l^{\lambda\gamma}l^{\beta\epsilon}\delta^\delta_{~(\mu}\delta^\alpha_{~\nu)}
+2g^{\delta\gamma}l^{\beta\epsilon}l^\lambda_{~(\mu}\delta^\alpha_{~\nu)}
\nonumber\\
&\qquad\qquad\qquad
-\left\{2g^{\alpha\gamma}l^{\beta\delta}l^{\lambda\epsilon}
+g^{\delta\alpha}l^{\lambda\gamma}l^{\beta\epsilon}
+g^{\delta\gamma}l^{\beta\epsilon}l^{\lambda\alpha}
\right\} g_{\mu\nu}\biggr) 
\partial_\alpha f_{\beta\gamma} =0,
\end{eqnarray}
where
\begin{eqnarray}
&N_{\delta\epsilon\lambda}:=M_{\delta\epsilon\lambda}+e_{a[\delta}\partial_{\epsilon]}f^a_{~\lambda},
\end{eqnarray}
and the brackets $(\cdot\cdot)$ denotes symmetrization, i.e.,
\begin{eqnarray}
T_{\alpha \cdots (\beta \gamma)\cdots} := \frac{1}{2}\left(
T_{\alpha \cdots \beta \gamma\cdots}+T_{\alpha \cdots \gamma\beta\cdots}\right),
\end{eqnarray}
and $\bar  F_{\mu\nu}$ is a function of $f_{\alpha\beta}$, $M_{\mu\nu\lambda}$, $f^a_{~\alpha}$, 
$\partial_\mu f^a_{~\alpha}$ and $\partial_\mu \partial_\nu f^a_{~\alpha}$. 
Note that derivatives of the dynamical variables $f_{\alpha\beta}$ and $M_{\mu\nu\lambda}$ do not appear in $\bar  F_{\mu\nu}$ and that derivatives of $f^a_{~\alpha}$ are not dynamical but fixed variables because $f^a_{~\alpha}$ is the given background tetrad. The constraint equations (\ref{vconst}) and (\ref{sconst}) can be written without derivatives of the dynamical variables $f_{\alpha\beta}$ and $M_{\mu\nu\lambda}$ as 
\begin{eqnarray}
&&
l^\nu_{~\mu}{\cal C}_\nu=2 l^{\alpha\beta} N_{\mu\alpha\beta}
+2l^\beta_{~\mu}e_a^{~\alpha}\partial_{[\alpha}f^a_{~\beta]}=0,  \label{cons}\\
&&{\cal B}:= {\cal C}-{\cal C}_{\mu}\bar g^{\mu\nu}{\cal C}_{\nu}\nonumber\\
&&\quad
= 4\alpha_0+3 \alpha_1 f
-g^{\mu\alpha}(2l^{\beta\lambda}l^{\gamma\nu}+g^{\beta\nu}\bar g^{\lambda\gamma})
N_{\mu\nu\lambda}N_{\alpha\beta\gamma} 
\nonumber\\
&&\qquad
+e^{~\mu}_{a}e^{~\nu}_{b}R(f)_{\mu\nu}^{~~ab}-2N_{\nu\rho\gamma}\left(2l^{\gamma\mu}l^{\beta\nu}e^{b\rho}{\cal F}_{\mu\beta b}
+e^{b\gamma}l^{\beta\nu}l^{\rho\lambda}{\cal F}_{\lambda\beta b}\right)
\nonumber\\
&&\qquad
+\left(2\eta^{ab}l^{\beta\nu}l^{\mu\alpha}
+ \bar g^{\nu\alpha}f^{a}_{~\lambda}e^{b\lambda}l^{\beta\mu}
\right){\cal F}_{\mu\nu a}{\cal F}_{\alpha\beta b}, \label{const}
\end{eqnarray}
with
\begin{eqnarray}
{\cal F}_{\alpha\beta }^{~~~c}:= \partial_{[\alpha} f^c_{~\beta]},
\end{eqnarray}
where, in accord with Ref.\cite{DW}, we use the combination (\ref{const}) of the constraint equations instead of ${\cal C}$. Here $R(f)_{\mu\nu}^{~~ab}$ is the Riemann tensor corresponding to the background vierbein $f^a_{~\mu}$.
Acted upon by the differential operator $\partial_0$, 
the two constraint equations (\ref{cons}) and (\ref{const}) now become time evolution equations.

Now, we have the equations (\ref{Gmunu}), $\partial_0 l^{\alpha}_{~\beta}{\cal C}_\alpha$ and $\partial_0 {\cal B}$ which are expected to describe the time evolution of $f_{00}$, $M_{0i0}$ and $M_{0ij}$. 
These equations are quasi-linear\footnote{Quasilinear means the coefficient of the highest-order derivatives only depend on derivatives with \emph{strictly} lower order.} for the first order derivatives of $f_{\mu\nu}$ and $M_{\mu\nu\lambda}$. 
Using the time-evolution equations for $f_{0i}$, $f_{ij}$, $M_{ij0}$ and $M_{ijk}$, i.e., Eqs.(\ref{0i0}), (\ref{0ij}), (\ref{0ij0}) and (\ref{0ijk}), 
we can eliminate their time derivatives from Eq.(\ref{Gmunu}), $\partial_0 l^{\alpha}_{~\beta} {\cal C}_\alpha$ and $\partial_0 {\cal B}$, 
and consequently only the time derivatives of $f_{00}$, $M_{0i0}$ and $M_{0ij}$ appear in these equations. 
The forms of the equations become 
\begin{eqnarray}
&E_{\mu\nu}:= 
2\left(l^{00}\delta^0_{~(\mu}\delta^i_{~\nu)}-g^{00}l^0_{~(\mu}\delta^i_{~\nu)}
-l^{0i}\delta^0_{~(\mu}\delta^0_{~\nu)}+g^{00}l^{0i}g_{\mu\nu}\right)\partial_0 M_{0i0}
\nonumber\\
&\qquad\qquad
+2\left(l^{0j}\delta^0_{~(\mu}\delta^i_{~\nu)}-g^{00}l^j_{~(\mu}\delta^i_{~\nu)}
-l^{ij}\delta^0_{~(\mu}\delta^0_{~\nu)}+g^{00}l^{ij}g_{\mu\nu}\right)\partial_0 M_{0ij}
\nonumber\\
&\qquad\qquad
+\biggl[2N_{\alpha\beta\gamma}\biggl(-l^{0\alpha}l^{0\gamma}\delta^\beta_{~(\mu}\delta^0_{~\nu)}
+g^{00}l^{0\alpha}\delta^\beta_{~(\mu}l^\gamma_{~\nu)}
+g^{0\alpha}l^{0\beta}l^\gamma_{~(\mu}\delta^0_{~\nu)} \nonumber\\
&\qquad\qquad\qquad\qquad\quad
-l^{\alpha\gamma}l^{0\beta}\delta^0_{~\mu}\delta^0_{\nu}
-g^{0\alpha}l^{0\beta}l^{0\gamma}g_{\mu\nu}
+g^{00}l^{\alpha\gamma}l^{0\beta}g_{\mu\nu}\biggr)  \nonumber\\
&\qquad\qquad\qquad\quad
+l^0_{~b}\left(\partial_\delta f^b_{~\lambda}\right)
\biggl(l^{\lambda0}\delta^0_{~(\mu}\delta^\delta_{~\nu)}
-l^{\lambda\delta}\delta^0_{~\mu}\delta^0_{~\nu}
+g^{0\delta}l^\lambda_{~(\mu}\delta^0_{~\nu)}
\nonumber\\
&\qquad\qquad\qquad\qquad
-g^{00}l^\lambda_{~(\mu}\delta^\delta_{~\nu)}
-g^{0\delta}l^{\lambda0}g_{\mu\nu}
+g^{00}l^{\lambda\delta}g_{\mu\nu} \biggr)
\biggr]\partial_0 f_{00} \nonumber\\
&\qquad
=F^{i\alpha\beta}_{\mu\nu} \partial_i f_{\alpha\beta}
+F^{i\alpha\beta\gamma}_{\mu\nu} \partial_i M_{\alpha\beta\gamma} 
+F_{\mu\nu},
\label{C1}\\
&E_{\mu}:=
2\left( l^{0i}\delta^0_{~\mu}-l^{00}\delta^i_{~\mu}\right)\partial_0 M_{0i0}
+2\left(l^{ij}\delta^0_{~\mu}-l^{0j}\delta^i_{~\mu}\right)\partial_0 M_{0ij}\nonumber\\
&\qquad
+2\biggl(-l^{0\gamma}l^{0\delta}N_{\mu\gamma\delta}
+\delta^0_{~\mu}l^0_{~\lambda}l^{\delta \lambda} e_a^{~\gamma}\partial_{[\gamma}f^a_{~\delta]}\nonumber\\
&\qquad\qquad\qquad\qquad
-l^{0\gamma}e_a^{~0}l^\delta_{~\mu}\partial_{[\gamma}f^a_{~\delta]}
+l^0_{~a}l^{\gamma\delta}\delta^0_{~[\mu}\partial_{\gamma]}f^a_{~\delta}
\biggr)\partial_0 f_{00}  \nonumber\\
&\qquad
=G^{i\alpha\beta}_{\mu} \partial_i f_{\alpha\beta}
+G^{i\alpha\beta\gamma}_{\mu} \partial_i M_{\alpha\beta\gamma}
+G_{\mu} ,
\label{C2}\\ 
&E:=
4\left( g^{i\alpha}l^{0\beta}l^{0\gamma} -g^{0\alpha}l^{0\beta}l^{i\gamma} 
-g^{0\alpha}g^{i\beta}\bar g^{0\gamma}\right)N_{\alpha\beta\gamma}\partial_0M_{0i0}
\nonumber\\
&\qquad\quad
+4\left( g^{i\alpha}l^{j\beta}l^{0\gamma} -g^{0\alpha}l^{j\beta}l^{i\gamma} 
-g^{0\alpha}g^{i\beta}\bar g^{j\gamma}\right)N_{\alpha\beta\gamma}\partial_0M_{0ij}
+\Xi \partial_0 f_{00}
\nonumber\\
&\qquad
=H^{i\alpha\beta}\partial_i f_{\alpha\beta}
+H^{i\alpha\beta\gamma} \partial_i M_{\alpha\beta\gamma}
+H, \label{C3}\\
\end{eqnarray}
with
\begin{eqnarray}
&\Xi:=-3\alpha_1 g^{00} 
-2l^{\mu 0} e_a^{~0}e_b^{~\nu}R_{\mu\nu}^{~~ab}
\nonumber\\
&\qquad
+4N_{\mu\nu\lambda}N_{\alpha\beta\gamma}
\left(g^{0\mu}l^{0\alpha}l^{\nu\gamma}l^{\lambda\beta}
+g^{\mu\alpha}l^{0\nu}l^{\lambda\beta}l^{0\gamma}
+g^{\mu\alpha}g^{0\nu}l^{0\beta}\bar g^{\gamma\lambda} \right) 
\nonumber\\
&\qquad
-2\left(g^{\mu  0}l^{\beta\lambda}l^{\gamma\nu}+g^{\mu  0}g^{\beta\nu}\bar g^{\lambda\gamma}
-g^{\mu  \beta}l^{0\lambda}l^{\gamma\nu}\right)
N_{\mu\nu\lambda} l^{0}_{~a}\partial_\beta f^a_{~\gamma}
\nonumber\\
&\qquad
-4l^{0}_{~a}\partial_\rho f^a_{~\gamma}\biggl(
l^{\gamma\mu}l^{\beta0}e^{b\rho}
+e^{b\gamma}l^{\beta0}l^{\rho\mu}
-l^{\gamma\mu}l^{\beta\rho}e^{b0}
\biggr){\cal F}_{\mu\beta b}
\nonumber\\
&\qquad
+ 2N_{\nu\rho\gamma}
\biggl(2l^{\gamma0}l^{0\mu}l^{\beta\nu}e^{b\rho}
+2l^{\gamma\mu}l^{\beta0}l^{0\nu}e^{b\rho}
\nonumber\\
&\qquad\qquad\qquad\qquad
+l^{\gamma\mu}l^{\beta\nu}e^{b0}l^{0\rho}
+2l^{\beta0}l^{0\nu}l^{\rho\mu}e^{b\gamma}\biggr){\cal F}_{\mu\beta b}
\nonumber\\
&\qquad
-\biggl(2\eta^{ab}l^{\beta0}l^{\nu0}l^{\mu\alpha}+2\eta^{ab}l^{\beta\nu}l^{\mu0}l^{0\alpha}
\nonumber\\
&\qquad\qquad\qquad\qquad
+\bar g^{\nu\alpha}e^{a0}e^{b0} l^{\beta\mu}+ \bar g^{\nu\alpha}e^{a\lambda}f^b_{~\lambda}l^{\beta0}l^{\mu0}
\biggr){\cal F}_{\mu\nu a}{\cal F}_{\alpha\beta b},
\end{eqnarray}
where $F^{i\alpha\beta}_{\mu\nu}$,
$F^{i\alpha\beta\gamma}_{\mu\nu}$,
$F_{\mu\nu}$,
$G^{i\alpha\beta}_{\mu}$,
$G^{i\alpha\beta\gamma}_{\mu}$,
$G_{\mu} $,
$H^{i\alpha\beta}$,
$H^{i\alpha\beta\gamma}$ and 
$H$ are functions of $f_{\alpha\beta}$, $M_{\mu\nu\lambda}$, $f^a_{~\alpha}$, 
$\partial_\mu f^a_{~\alpha}$ and $\partial_\mu \partial_\nu f^a_{~\alpha}$, and contain no derivative terms of  $f_{\alpha\beta}$ and $M_{\mu\nu\lambda}$. 
Although Eqs.(\ref{C1}), (\ref{C2}) and (\ref{C3}) seem to have fifteen components in total, only ten of the fifteen are actually independent time-evolution equations.
This can be seen as follows:
For $\mu=0$ in Eq.(\ref{C1}), the left hand side becomes zero, and thus it does not give any time evolution. 
Moreover, the trace of the left hand side of Eq.(\ref{C1}) is proportional to the left hand side of Eq.(\ref{C2}) with $\mu=0$, which means that the trace does not give any independent information for the time evolution. 
This is consistent with the discussion in the previous section,
i.e., five of ten degrees of freedom of Eq.(\ref{EOM}) should become constraints as in Eq.(\ref{vconst}) and Eq.(\ref{sconst}). 

To conclude, we now have ten time-evolution equations. 
We can finally write down the characteristic equations. 
They are:
\begin{eqnarray}
&\frac{1}{4} \Xi \tilde f_{00}
+ \sum_i \left( g^{i\alpha}l^{0\beta}l^{0\gamma} -g^{0\alpha}l^{0\beta}l^{i\gamma} 
-g^{0\alpha}g^{i\beta}\bar g^{0\gamma}\right)N_{\alpha\beta\gamma} \tilde M_{0i0}
\nonumber\\
&\qquad
+\sum_i \left( g^{i\alpha}l^{i\beta}l^{0\gamma} -g^{0\alpha}l^{i\beta}l^{i\gamma} 
-g^{0\alpha}g^{i\beta}\bar g^{i\gamma}\right)N_{\alpha\beta\gamma} \tilde M_{0ii}
\nonumber\\
&\qquad
+\sum_{i>j}2\left( g^{i\alpha}l^{j\beta}l^{0\gamma} -g^{0\alpha}l^{j\beta}l^{i\gamma} 
-g^{0\alpha}g^{i\beta}\bar g^{j\gamma}\right)N_{\alpha\beta\gamma}\tilde M_{0(ij)} =0,\\
\nonumber\\
& l^{00} \tilde M_{0i0} +  l^{0i} \tilde M_{0ii} + \sum_{i\neq j} l^{0j} \tilde M_{0(ij)} 
\nonumber\\
&\qquad
+\biggl(l^{0\alpha}l^{0\beta}N_{i\alpha \beta} 
-l^{0\gamma}e_a^{~0}l^\delta_{~i}\partial_{[\gamma}f^a_{~\delta]}
+\frac{1}{2}l^0_{~a}l^{0\delta}\partial_{i}f^a_{~\delta}
\biggr)\tilde f_{00}=0, \\
\nonumber\\
&\sum_k \left(l^0_{~(i}\delta^k_{j)}-l^{0k}g_{ij}\right) \tilde M_{0k0}
+\sum_k \left(l^k_{~(i}\delta^k_{j)}-l^{kk}g_{ij}\right) \tilde M_{0kk}
\nonumber\\
&\qquad
+2\sum_{k>l} \left(l^k_{~(i}\delta^l_{j)}-l^{lk}g_{ij}\right) 
\tilde M_{0(kl)} \nonumber\\
&\qquad
+\biggl[N_{\mu\nu\lambda}\left(-l^{0\mu}\delta^\nu_{~(i}l^\lambda_{~j)}
-g^{0\mu}l^{0\nu}l^{0\lambda}g_{ij} -l^{\mu\lambda}l^{0\nu}g_{ij}\right)
\nonumber\\
&\qquad\qquad
+\frac{1}{2}l^0_{~b}\left(\partial_\delta f^b_{~\lambda}\right)
\biggl(
l^\lambda_{~(i}\delta^\delta_{~j)}
-g^{0\delta}l^{\lambda0}g_{ij}
-l^{\lambda\delta}g_{ij} \biggr)
\biggr]
 \tilde f_{00} =0, 
\end{eqnarray}
where we explicitly show the summation with respect to spatial coordinates because some of the indices do not obey Einstein's summation rule. 
Here, we decompose $M_{0ij}$ into $M_{0(ij)}$ and $M_{0[ij]}$, but $M_{0[ij]}$ does not appear in the characteristic equations because from Eq.(\ref{constM}) we see that $M_{0[ij]}$ can be expressed in terms of $M_{ij0}$.
Here, the notation $\tilde M_{\mu\nu\lambda}$ and $\tilde f_{00}$ means they are not the values of $M_{\mu\nu\lambda}$ and $f_{00}$, but represent the ``(finite) difference'' of the fields in the $x^0$-direction. 
The term ``difference" is related to the $x_0$-derivative in the case when the derivative is well-defined. In the context of calculus of single variable $y=y(x)$, the ``difference'' is the quantity $\Delta y/\Delta x$, related to the derivative $dy/dx$ by the well-known definition $dy/dx=\lim_{\Delta x\to 0} \Delta y/\Delta x$.
Here we do not use the $x_0$-derivative but ``difference" because, in characteristic analysis, sometimes we consider shock waves on which derivatives are not well-defined 
(See the discussion in \ref{app}).

\section{Characteristic Equations with Non-Vanishing Contortion}
\label{section4}

In Ref.\cite{DW}, it was argued that, if the contortion 
does not vanish, an arbitrary hypersurface can be a characteristic hypersurface. 
If so, a generic solution always has acausal structure, and thus the theory is very problematic. 
With our characteristic equations, we revisit this problem.  

We first consider the simplest example with the Minkowski background tetrad. 
Suppose the dynamical tetrad completely coincides with the background tetrad, i.e., 
$f^a_{~\mu}=e^a_{~\mu}$, 
and let us examine the characteristic determinant corresponding to a flat spatial hypersurface. 
Under this situation, the expressions of $g_{\mu\nu}$, $\bar g_{\mu\nu}$, $f_{\mu\nu}$, $l_{\mu\nu}$, $M_{\mu\nu\lambda}$ and $N_{\mu\nu\lambda}$ become
\begin{eqnarray}
g_{\mu\nu}=\bar g_{\mu\nu}=f_{\mu\nu}=l_{\mu\nu}=\eta_{\mu\nu}, \quad \mbox{and} \quad
M_{\mu\nu\lambda}=N_{\mu\nu\lambda}=0.
\end{eqnarray}
Then the characteristic equations simplify significantly and reduce to
\begin{eqnarray}
&&-\tilde M_{011}-\tilde M_{022}=0, \quad 2\tilde M_{0(12)}=0, \quad -\tilde M_{010}=0,
\nonumber\\
&&-\tilde M_{022}-\tilde M_{033}=0, \quad 2\tilde M_{0(23)}=0, \quad -\tilde M_{020}=0,
\nonumber\\
&&-\tilde M_{033}-\tilde M_{011}=0, \quad 2\tilde M_{0(31)}=0, \quad -\tilde M_{030}=0,
\quad -\frac{3}{4}\alpha_1 \tilde f_{00}=0.
\end{eqnarray}
The characteristic determinant of this system is obviously generically nonzero and its eigenvalues are of order $O(1)$ 
and $O(\alpha_1)$, which are finite constants. 

Now we consider a tiny perturbation on Minkowski space and consider the changes in the characteristic determinant. 
Since the characteristic determinant on Minkowski space is finite and nonzero, by continuity, with a sufficiently small perturbation
the characteristic determinant must still be nonzero \emph{even if the contortion does not vanish}. This means that the hypersurface is \emph{not} a characteristic hypersurface (although for some values of the fields, the characteristic matrix could become zero and thus a superluminal mode could potentially arise. We comment on this further in section \ref{section5}). This is in conflict with the conclusion of Ref.\cite{DW}. There is thus a need to point out why our analysis gives a result that is different from that of Ref.\cite{DW}.
 
In our analysis, the first order derivative of $f_{00}$ joins in the characteristic equations, while in the analysis of Ref.\cite{DW} it does not. 
The reason why it does not appear in their analysis is that the authors assume all discontinuities across a shock surface only show up in the second order derivative of the dynamical metric. However, as we will show in Appendix A, in the discussion of characteristic we should define the characteristic matrix such that every dynamical field appears in the general solutions. Indeed, except for the contribution from $\partial_0 f_{00}$, our result is otherwise consistent with that of Ref.\cite{DW}. 
However, since their characteristic matrix does not have the contribution from $\partial_0 f_{00}$, its determinant always becomes zero. 
If we go back to the original viewpoint of the Cauchy-Kovalevskaya theorem and the discussion about shock waves, 
it is reasonable for $\partial_0 f_{00}$ to appear in characteristic matrix. This is further discussed in Appendix A. 

\section{Discussion: Does Non-Linear Massive Gravity Admit Superluminal Propagation?}
\label{section5}

We have carefully examined the characteristic equations of a special case of non-linear massive gravity and showed that Cauchy problem can be well-posed, in the sense that not every hypersurface can be a characteristic hypersurface given a physical metric. 
In the viewpoint on the fixed fiducial metric, of course, on an arbitrary hypersurface there probably exists specific field values, i.e., the specific physical metrics, which give vanishing characteristic determinant. In this sense, any hypersurface potentially \emph{can} be characteristic by tuning values of a physical metric.
However, the situation here is a bit more subtle than the   
result in Ref.\cite{DW}, in which the authors argued that, any hypersurface can be characteristic on a given physical metric. The difference is due to the contribution of the first derivative term of $f_{00}$, which should appear in the characteristic equation in view of the discussion in \ref{app}.  

We emphasize again that our result does \emph{not} guarantee that the theory is free of problems.
Indeed, we have merely shown that there exist good non-characteristic hypersurfaces, \emph{not} the \emph{absence} of superluminal characteristics. One should next examine the characteristic matrix carefully to determine if its determinant admits any real root, the presence of which would imply the existence of superluminal propagations.
The existence of such real root will directly affect the discussion of causality. 
It reduces the Cauchy region and is sometimes responsible for giving rise to acausality.
In order to have a deeper understanding of the causal structure of the theory,
we will need an elaborate inspection of the characteristics. We shall leave this issue for future work. Nevertheless, we would like to emphasize that superluminality does \emph{not} always imply acausality \cite{Bruneton, Afshordi, Geroch1, BCG}. In addition, one should check if occurrence of superluminal modes simply signal that the theory as an \emph{effective} theory is breaking down, \emph{\`a la} Ref.\cite{BRHT}. 

It is probably a good guess that non-linear massive gravity does suffer from superluminal propagation in general and that this might result in acausality.
Indeed, we have seen tantalizing hints of superluminal propagation on the self-accelerating solution \cite{Shinji}, and in fact, \emph{instantaneous} propagation may arise in the full theory with all the $\alpha_n$'s being nonzero \cite{Chien-I}. In particular, energy can probably be emitted with infinite speed on the self-accelerating background by the helicity-0 mode of the massive graviton. In addition to superluminal mode, non-linear massive gravity on the self-accelerating solution has the property that although it generically has 5 degrees of freedom \cite{HR}, in the second order action on open FLRW background, perturbative analysis only reveals two tensors degrees of freedom \cite{GLM}. That is, the full theory has different number of degrees of freedom compared to some of the low order limits. This could be dangerous as the excitations of the extra degrees of freedom may be accompanied by anomalous characteristic, as discussed in the context of $f(T)$ gravity \cite{OINC} and Poincar\'e gauge theory \cite{Nester8}. 
Solutions with such anomalous characteristic is inappropriate to describe our Universe. Furthermore non-linear instability has also been pointed out in the self-accelerating solution \cite{DeFelice:2012mx}. 

There remain a few puzzles to be resolved. Firstly, a further careful analysis is required to check if non-linear massive gravity with $\left\{\alpha_n\right\}_{n=2,3,4}=0$ does admit superluminal propagation as argued in Ref.\cite{DW}, which might well be the case. Secondly, it would be interesting to investigate how the superluminal helicity-0 mode arises in the full theory and its corresponding characteristic analysis. 
Solving them, we can assess the appropriateness of each solution, and it might give us further constraints on the theory.

\ack

The authors would like to thank James Nester for valuable discussions and helpful comments that helped improved this work considerably. The authors are also grateful to 
Pisin Chen for much appreciated help and various supports. Keisuke Izumi is supported by Taiwan National Science Council under Project No. NSC101-2811-M-002-103. Yen Chin Ong is supported by the Taiwan Scholarship from Taiwan's Ministry of Education.

\appendix

\section{Review of characteristic Analysis \emph{\`a la} Cauchy and Kovalevskaya}
\label{app}

The analysis of the characteristic is related to the Cauchy-Kovalevskaya theorem, from which the propagation of shock-wave front can be obtained. 
Our discussion here is based on the equations that appear in the usual proof of the Cauchy-Kovalevskaya theorem (see, e.g., the seminal work of Courant and Hilbert \cite{CH}). 
Since the naive extension of the result in the case involving only a single field to that of a multi-field case is invalid, 
we re-visit the original arguments used in the proof of the Cauchy-Kovalevskaya theorem and derive the exact characteristic equations in the multi-field case.

\subsection{Single Field}
Suppose we have an $n$-th order differential equation which is quasilinear for the $n$-th order derivative terms, i.e. 
\begin{equation}
A^{\alpha_1\cdots\alpha_n}\partial_{\alpha_1\cdots\alpha_n} \phi
+A=0,
\end{equation}
where $\phi$ is a field and $A^{\alpha_1\cdots\alpha_n}$ and $A$ are functions of $x^\mu$ and derivatives of $\phi$ up to the $(n-1)$-th order.
We now introduce a hypersurface $\Sigma$ which is orthogonal to $\xi^\mu$. If 
\begin{eqnarray}
A^{\alpha_1\cdots\alpha_n}\xi_{\alpha_1}\cdots\xi_{\alpha_n}=0,
\label{single}
\end{eqnarray}
then the hypersurface $\Sigma$ is called a \emph{characteristic hypersurface}.
On the other hand, if 
\begin{eqnarray}
A^{\alpha_1\cdots\alpha_n}\xi_{\alpha_1}\cdots\xi_{\alpha_n}\neq 0,
\end{eqnarray}
the hypersurface $\Sigma$ is called a \emph{non-characteristic hypersurface}.
We have the following important result \cite{C,K}: 

\ 

\textsf{Cauchy-Kovalevskaya Theorem:}
\textit{If all the coefficients $A^{\alpha_1\cdots\alpha_n}$ and $A$ are analytic and 
if the hypersurface $\Sigma$ is a non-characteristic hypersurface, then
there exists a unique local analytic solution $\phi$ in the neighbourhood of $\Sigma$.}

\ 

This theorem implies that the uniqueness of time-evolution from a charactaristic hypersurface is not guaranteed. 
Moreover, there is an important fact about charactaristics, namely:

\ 

\textsf{Remark:}  \textit{A charactaristic hypersurface can be a shock-wave front.}

\ 

This remark means that there could be non-linear propagation on the characteristic hypersurface. 
It can be intuitively understood as follows:
Since we have the $n$-th order differential equation of motion for single field $\phi$, we can introduce, on the initial hypersurface $\Sigma_0$, $n$ numbers of initial conditions for $(\partial_t)^{n-1} \phi$,$\cdots$, $\partial_t\phi$ and $\phi$ where $\xi^\mu:= (\partial/ \partial t)^\mu$. 
If the initial hypersurface $\Sigma_0$ is not characteristic, we can derive $(\partial_t)^n \phi$ from the equation of motion, and then the values $(\partial_t)^{n-1} \phi$,$\cdots$, $\partial_t\phi$ and $\phi$ after a small time evolution from the initial hypersurface $\Sigma_0$ can be uniquely fixed as 
\begin{eqnarray}
(\partial_t)^{k-1} \phi (t_0+ \Delta t) =
(\partial_t)^{k-1} \phi (t_0) +
(\partial_t)^k \phi (t_0) \Delta t, \qquad (1 \le k\le n),
\end{eqnarray}
where $t=t_0$ on the initial hypersurface $\Sigma_0$.
However, if the hypersurface $\Sigma_0$ is characteristic, $(\partial_t)^n \phi$ cannot be fixed uniquely, which allows a discontinuity in $(\partial_t)^n \phi$ and thus non-unique time evolution of $(\partial_t)^{n-1} \phi$. 

In order to make the discussion in the next subsection clearer, 
we comment here that, in the usual proof of Cauchy-Kovalevskaya theorem \cite{CH}, a quasilinear $n$-th order differential equation for single field $\phi$ is reduced to a system of quasilinear first order differential equations for the multi-field ${\bf u}$, where 
\begin{eqnarray}
{\bf u} := \left( \phi, \frac{\partial \phi}{\partial x^0}, \cdots, \frac{\partial \phi}{\partial x^d}, \frac{\partial^2 \phi}{\partial x^0\partial x^0}, \cdots\right),
\label{singleu} 
\end{eqnarray}
and ${\bf u}$ includes all partial derivatives of $\phi$ up to the $(n-1)$-th order. 
Let $m$ denote the number of components of ${\bf u}$.
Then, Eq.(\ref{single}) can be exactly represented by 
\begin{eqnarray}
B^t\left[{\bf u} ,x \right]\partial_t {\bf u} = B^i \left[{\bf u} ,x \right]\partial_i {\bf u}
+B \left[{\bf u} ,x \right], 
\label{B}
\end{eqnarray}
where $B^t\left[{\bf u} ,x \right]$, $B^i\left[{\bf u} ,x \right]$ and $B \left[{\bf u} ,x \right]$ depend on ${\bf u}$ and $x$. The first two are $m\times m$ matrices and the last one is a $m$-vector.
Time evolution of every component in ${\bf u}$ can be obtained if $\det B^t\neq 0$. 
While most of the eigenvalues of $B^t\left[{\bf u} ,x \right]$ are always nonzero, the only nontrivial eigenvalue is the same as $A^{t\cdots t}$ in Eq.(\ref{single}). 
In the proof of the Cauchy-Kovalevskaya theorem, after transforming into Eq.(\ref{singleu}),
the uniqueness is proved under the condition $\det B^t\neq 0$ (Of course we also need the analyticity condition). 

To conclude, in the Cauchy-Kovalevskaya theorem and in the discussion of shock waves, the only important term  is $A^{t\cdots t}$. If $A^{t\cdots t}=0$, it is outside the range of applicability of the Cauchy-Kovalevskaya theorem to ensure a well-posed Cauchy problem, and a shock wave can indeed propagate on the characteristic hypersurface.

\subsection{Multi-Field}

Here, we extend the discussion in the previous subsection to a multi-field case. While this can sometimes be extended straightforwardly, it is in general, quite nontrivial. 
We will give examples of both cases in this subsection. 

\subsubsection{The Simple Case}

We consider equations for two fields $\phi$ and $\psi$,
\begin{eqnarray}
&\!\!\!\!\!\!
S^{\mu\nu}\partial_\mu\partial_\nu \phi
+T^{\mu\nu}\partial_\mu\partial_\nu \psi 
+ S =0, \\
&\!\!\!\!\!\!
U^{\mu\nu}\partial_\mu\partial_\nu \phi
+V^{\mu\nu}\partial_\mu\partial_\nu \psi
+ U =0,
\end{eqnarray}
where $S^{\mu\nu}, T^{\mu\nu}, S, U^{\mu\nu}, V^{\mu\nu}$ and $U$ are all functions of 
$\partial_\alpha \phi$, $\phi$, $\partial_\beta \psi$, $\psi$ and $x$. 
Picking up the higher order derivative terms with respect to $t$, 
we have
\begin{eqnarray}
{\bf M} \left(
  \begin{array}{c}
    \tilde \phi   \\
    \tilde \psi  \\
  \end{array}
\right) =0 ,
\qquad \mbox{with} \qquad
{\bf M}:=\left(
  \begin{array}{cc}
   S^{tt}    & T^{tt}   \\
    U^{tt}   &  V^{tt}  \\
  \end{array}
\right),
\end{eqnarray}
where $\tilde \phi$ and $\tilde \psi$ mean they are not the values of $\phi$ and $\psi$, but represent the ``difference" of the fields in the $t$-direction. 
If $\det {\bf M} =0$, we cannot derive one of the linear combinations of $\tilde \phi$ and $\tilde \psi$, and the hypersurface is a characteristic hypersurface, according to the discussion in the previous subsection.

\subsubsection{The Nontrivial Case}

Here, we consider the equations for two fields $\zeta$ and $\sigma$,
\begin{eqnarray}
&&I^{\mu\nu}\partial_\mu\partial_\nu\zeta
+J^{\mu}\partial_\mu \sigma
+I =0,\label{IJ}\\
&&L =0, \label{L}
\end{eqnarray}
where $I^{\mu\nu}$, $J^{\mu\nu}$, $I$ 
and $L$ are all functions of $\partial_\alpha \zeta$, $\zeta$, $\sigma$ and $x$. 
Actually, the equations in the non-linear massive gravity theory are very similar to these equations. 
To obtain the second order differential equation, we operate with $\partial_\mu$ on Eq.(\ref{L}) and obtain
\begin{eqnarray}
\frac{\partial L}{\partial \left(\partial_\alpha \zeta\right)}\partial_\mu \partial_\alpha \zeta +\frac{\partial L}{\partial \zeta}\partial_\mu  \zeta +\frac{\partial L}{\partial \sigma}\partial_\mu  \sigma +\frac{\partial L}{\partial x^\mu}=0.\label{pL}
\end{eqnarray}
If we naively pick up the higher-order derivative terms, the characteristic equations are
\begin{eqnarray}
{\bf \bar N} \left(
  \begin{array}{c}
    \tilde \zeta   \\
    \tilde \sigma  \\
  \end{array}
\right) =0 ,
\qquad \mbox{with} \qquad
{\bf \bar N}:=\left(
  \begin{array}{cc}
   I^{tt}    & 0   \\
    \frac{\partial L}{\partial \left(\partial_t \zeta\right)}   & 0  \\
  \end{array}
\right), \label{barN1}
\end{eqnarray}
where $\tilde \zeta$ and $\tilde \sigma$ mean they are not the values of $\zeta$ and $\sigma$, but represent the ``difference" of the fields in the $t$-direction. 
Then, $\det {\bf \bar N}$ is trivially zero. 
Does it mean that every hypersuface is a characteristic hypersurface? 

In order to investigate this question, first we consider a simple example
\begin{eqnarray}
L=P +Q \sigma, \label{Q}
\end{eqnarray}
where $P$ is a function of $\partial_\alpha \zeta$ and $\zeta$, while $Q$ is a non-zero constant.
Then, Eq.(\ref{L}) can be regarded as a constraint equation for fixing the auxiliary field $\sigma$ and we can solve it for $\sigma$. 
Substituting the solution into Eq.(\ref{IJ}), we can obtain an equation for $\zeta$:
\begin{eqnarray}
\left(QI^{\mu\nu}+ J^{\mu}\frac{\partial P}{\partial (\partial_\nu\zeta)}\right)\partial_\mu\partial_\nu\zeta
+\bar I  =0, 
\label{zeta}
\end{eqnarray}
where $\bar I$ is  a function of $\partial_\alpha \zeta$, $\zeta$ and $x$. 
Now, Eq.(\ref{zeta}) is the equation for the single field $\zeta$ and 
the charactaristic equation for $\zeta$ is obviously
\begin{eqnarray}
\left(QI^{tt}+ J^{t}\frac{\partial P}{\partial (\partial_t\zeta)} \right) \tilde \zeta=0,
\label{QI}
\end{eqnarray}
which is different from Eq.(\ref{barN1}). 
From the discussion in the previous subsection, $\zeta$ is uniquely fixed in the neighborhood if
\begin{eqnarray}
QI^{tt}+ J^{t}\frac{\partial P}{\partial (\partial_t\zeta)} \neq 0,
\end{eqnarray}
and then, $\sigma$ can be also determined uniquely from Eqs.(\ref{L}) and (\ref{Q}).
Therefore, we find that Eq.(\ref{QI}) is the exact characteristic equation instead of Eq.(\ref{barN1}). 

How do we derive the characteristic equation in a general multi-field case? 
Going back to Eqs.(\ref{IJ}) and (\ref{L}), it can be seen that in both equations, the order of derivative operators of $\sigma$ are one less than those of $\zeta$. 
This means that discontinuity of $\left(\partial_t\right)^k \zeta$ can be balanced with that of $\left(\partial_t\right)^{k-1} \sigma$.
Therefore, not only the second order derivative of $\zeta$ but also the \emph{first} order derivative of $\sigma$ should appear in the characteristic equation. 
Then, the characteristic equation can be written as 
\begin{eqnarray}
{\bf N} \left(
  \begin{array}{c}
    \tilde \zeta   \\
    \tilde \sigma  \\
  \end{array}
\right) =0 ,
\qquad \mbox{with} \qquad
{\bf N}:=\left(
  \begin{array}{cc}
   I^{tt}    & J^t   \\
    \frac{\partial L}{\partial(\partial_t \zeta)}   & \frac{\partial L}{\partial \sigma}  \\
  \end{array}
\right), \label{barN}
\end{eqnarray}
and the condition of the characteristic is consistent with Eq.(\ref{QI}). 

This statement can be justified further by writing the equations in the form of Eq.(\ref{B}). Let
us introduce new variables 
\begin{eqnarray}
&&v_i:=\partial_i \zeta\label{v},\\
&&u:=\partial_t \zeta,\label{u}
\end{eqnarray}
which are related by the commutativity of partial derivatives:
\begin{eqnarray}
\partial_t v_i = \partial_i u.\label{uv}
\end{eqnarray}
Eqs.(\ref{IJ}) and  (\ref{pL}) can be represented by
\begin{eqnarray}
&&I^{tt}\partial_t u+J^{t}\partial_t \sigma=-\left(I^{ti}+I^{it}\right)\partial_i u -I^{ij}\partial_i v_j
-J^{i}\partial_i \sigma- I,\label{chara1}\\
&&
\frac{\partial L}{\partial \left(\partial_t \zeta\right)}\partial_t u
+\frac{\partial L}{\partial \sigma}\partial_t  \sigma
=-\frac{\partial L}{\partial \left(\partial_i \zeta\right)}\partial_i u
-\frac{\partial L}{\partial \zeta}\partial_i  \zeta 
-\frac{\partial L}{\partial x^t}\label{chara2},
\end{eqnarray}
where all coefficients $I^{\mu\nu}$, $J^\mu$, $I$, ${\partial L}/{\partial \left(\partial_\mu \zeta\right)}$, ${\partial L}/{\partial  \zeta}$, ${\partial L}/{\partial  \sigma}$ and ${\partial L}/{\partial x^t}$ are written in non-derivative variables $\zeta$, $\sigma$, $u$ and $v_i$. 
The combination of Eqs.(\ref{u})--(\ref{chara2}) is exactly of the form of Eq.(\ref{B}). 
We can then apply the Cauchy-Kovalevskaya theorem and also have the same discussion about shock waves.
Again, we can deduce that $\det {\bf N}=0$ is the condition for the hypersurface to be the characteristic hypersurface.

Finally, we discuss in brief the method for deriving characteristic equations with general multi-fields. 
We start by choosing one of the fields and define the dimension of it as zero. The dimension, denoted by $[\cdot]$, of the differential operator is one, i.e. $[\partial]=1$. We determine the dimensions of other fields such that each field necessarily appears in one of the characteristic equations. 
In the case of Eqs.(\ref{IJ}) and (\ref{L}), for instance, we define the dimension of $\zeta$ as zero. If the dimension of $\sigma$ is larger than one, only $\partial_\mu \sigma$ can appear in the characteristic equations. 
On the other hand, if the dimension of $\sigma$ is smaller than one, only $\partial_\mu \partial_\nu \zeta$ can appear in the characteristic equations. Therefore, only in the case where the dimension of $\sigma$ is equal to one, both can appear in the characteristic equations, in which case we have the situation discussed above.

\end{document}